\documentclass[12pt]{iopart}

\usepackage{iopams}  
\usepackage{graphicx}
\begin{document}

\title[Monoclinic SrIrO$_3$]{Monoclinic SrIrO$_3$　– A Dirac semimetal produced by non-symmorphic symmetry and spin-orbit coupling}

\author{T Takayama$^{1,2}$, A N Yaresko$^1$, and H Takagi$^{1,2,3}$}

\address{$^1$Max Planck Institute for Solid State Research, Heisenbergstrasse
  1, 70569 Stuttgart, Germany}
\address{$^2$Institute for Functional Matter and Quantum Technologies, University of Stuttgart, Pfaffenwaldring 57, 70550 Stuttgart, Germany}
\address{$^3$Department of Physics, University of Tokyo, 7-3-1 Hongo, Tokyo 113-0033, Japan}
\ead{t.takayama@fkf.mpg.de}
\vspace{10pt}

\begin{abstract}
SrIrO$_3$ crystallizes in a monoclinic structure of distorted hexagonal perovskite at ambient pressure. The transport measurements show that the monoclinic SrIrO$_3$ is a low-carrier density semimetal, as in the orthorhombic perovskite polymorph. The electronic structure calculation indicates a semimetallic band structure with Dirac bands at two high-symmetry points of Brillouin zone only when spin-orbit coupling is incorporated, suggesting that the semimetallic state is produced by the strong spin-orbit coupling. We argue that the Dirac bands are protected by the non-symmorphic symmetry of lattice.
\end{abstract}

%

%

%
%
%

\section{Introduction}
5$d$ transition-metal oxides recently emerged as a platform for the exploration of novel electronic states produced by the interplay of modest Coulomb interaction $U$ with strong spin-orbit coupling $\lambda_{\rm SO}$ \cite{Krempa2014}. Among the 5$d$-based oxides, complex iridium oxides with Ir$^{4+}$ ions (5$d^5$) have been intensively investigated. In many of them, the Ir$^{4+}$ ions are surrounded octahedrally by oxygen ions, and their five $d$-electrons are accommodated into the $t_{2g}$ manifold due to a large crystal field splitting. The strong spin-orbit coupling of Ir as large as 0.5 eV splits the $t_{2g}$ manifold into the lower-lying fully-filled $J_{\rm eff}$ = 3/2 quartet and the upper $J_{\rm eff}$ = 1/2 doublet with one electron (hole), forming a half-filled band of   predominant $J_{\rm eff}$ = 1/2 character \cite{Kim2009}. The $J_{\rm eff}$ = 1/2 wave function consists of equal superpositions of three $t_{2g}$ orbitals including complex component.

In the Ir$^{4+}$ oxides, even a moderate Coulomb $U$ of 1-2 eV can marginally open a charge gap in $J_{\rm eff}$ = 1/2 bands and a $J_{\rm eff}$ = 1/2 Mott insulating state is often realized. The magnetism of $J_{\rm eff}$ = 1/2 Mott insulator has been one of the central topics in the spin-orbital entangled 5$d$ systems \cite{Jackeli2009}. The spin-orbital entangled character of $J_{\rm eff}$ = 1/2 wave function gives rise to bond-sensitive exchange coupling and the resultant unconventional magnetic ground states, as demonstrated by the two-dimensional Heisenberg antiferromagnetism in the layered perovskite Sr$_2$IrO$_4$ \cite{Fujiyama2012} and the Kitaev magnetism in honeycomb-based iridates \cite{Winter2017}.

The $J_{\rm eff}$ = 1/2 Mott state, marginally stabilized by the modest Coulomb $U$, is not always the case for Ir$^{4+}$ oxides. Indeed, metallic iridates are obtained for Ir$^{4+}$ oxides with a relatively large $J_{\rm eff}$ = 1/2 bandwidth such as those with three-dimensional crystal structure. The examples include the pyrochlore iridate Pr$_2$Ir$_2$O$_7$ with the largest A-site ion among the A$_2$Ir$_2$O$_7$ family (A: trivalent cations) \cite{Nakatsuji2006}, and the three-dimensional perovskite SrIrO$_3$ \cite{Zhao2008}. One may naively expect that all of the $J_{\rm eff}$ = 1/2 metals have a large Fermi surface originating from the half-filled $J_{\rm eff}$ = 1/2 bands. These metallic iridates, however, were very often found to be a semimetal \cite{Kondo2015, Matsuno2015}, which is discussed to be a consequence of the interplay of strong spin-orbit coupling and the lattice distortion. 

The perovskite SrIrO$_3$ has a GdFeO$_3$-type orthorhombic structure inheriting tilting and rotation of IrO$_6$ octahedra \cite{Longo1971}. Because of the small ion size of A-cation (Sr$^{2+}$ ion) with respect to the B-site (Ir$^{4+}$ ion), the orthorhombic perovskite phase can be stabilized only under a high physical pressure or under an epitaxial strain in a thin-film form. The transport properties evidence that the orthorhombic SrIrO$_3$ is a semimetal with a small density of carriers \cite{Matsuno2015, Nie2015}. The electronic structure calculation, together with the tight-binding analysis, suggested that the $J_{\rm eff}$ = 1/2 bands form a semimetallic electronic structure comprising the Dirac bands with a line of nodes right below the Fermi energy \cite{Carter2012, Zeb2012}. The presence of linearly-dispersive bands was confirmed by the ARPES measurement \cite{Nie2015}. The Dirac line node of orthorhombic SrIrO$_3$ is argued to be protected by the non-symmorphic symmetry containing two glide-operations (Space group: $Pbnm$, No. 62) \cite{Chen2016}.

At ambient pressure, SrIrO$_3$ crystallizes in a monoclinic structure of distorted 6H-type hexagonal perovskite (Space group: $C2/c$, no. 15) illustrated in Fig. 1(a) \cite{Longo1971}, distinct from the orthorhombic GdFeO$_3$-type structure. There are two iridium crystallographic sites in this structure, and both Ir atoms are placed in the strongly distorted oxygen octahedra. The Ir(1)O$_6$ octahedra form a triangular lattice in the $ab$-planes and are connected with Ir(2)O$_6$ octahedra along the $c$-axis by sharing the corner oxygens. A pair of Ir(2)O$_6$ octahedra form a face-sharing Ir(2)$_2$O$_9$ unit along the $c$-axis and form a triangular lattice in the $ab$-plane. The Ir(1)Ir(2)O$_9$ bilayers consisting of the Ir(1)O$_6$ layer and one Ir(2)O$_6$ layer connected by the corner oxygens may be viewed as a buckled honeycomb structure as in the (111) superlattice of cubic perovskite \cite{Xiao2011}.

The monoclinic SrIrO$_3$ was reported to be a metal as in the orthorhombic phase, which was discussed to be a non-Fermi liquid based on the unconventional power-law behavior of resistivity and enhanced magnetic susceptibility \cite{Cao2007}. The origin of non-Fermi liquid-like behavior is not clarified yet. In this paper, we present evidences showing that the monoclinic SrIrO$_3$ is a spin-orbit coupling induced semimetal as its orthorhombic polymorph. The semimetallic state consists of the two gapless Dirac bands at the A and M points of Brillouin zone near the Fermi energy. We argue that the Dirac points are protected by the non-symmorphic symmetry.

\section{Experimental}

A polycrystalline sample of monoclinic SrIrO$_3$ was prepared by a conventional solid-state reaction. The powder x-ray diffraction pattern of the product indicated almost single phase of monoclinic SrIrO$_3$. The magnetic, transport and thermodynamic measurements were performed on the polycrystalline pellet with using commercial instruments (Quantum Design, MPMS and PPMS). The thermoelectric properties such as Seebeck and Nernst constants were measured by the home-made apparatus. 

The electronic structure calculation was performed based on the local density approximation (LDA) with using the fully relativistic linear muffin-tin orbital (LMTO) method implemented in the PY LMTO code \cite{Antonov2004}. Spin-orbit coupling was incorporated by solving the four-component Dirac equation inside an atomic sphere. The effect of spin-orbit coupling is clarified by calculating the $J$-resolved density of states (DOS) where $J = l \pm 1/2$ is the total angular momentum for the 5$d$ states.

\section{Results}
\subsection{Transport and thermodynamic properties}

The resistivity measurement confirmed that the monoclinic SrIrO$_3$ is a metal in agreement with the previous report (Fig. 2(a)) \cite{Cao2007}. The magnetic susceptibility $\chi(T)$ in Fig. 1(b) shows an almost temperature-independent behavior, with a Curie-like upturn at low temperatures likely originating from impurities or defects. By subtracting the Curie-like contribution, the $T$-independent magnetic susceptibility $\chi_0$ was estimated to be $\sim$1.5 $\times$ 10$^{-4}$ emu/mol. A moderately large electronic specific heat coefficient $\gamma$ $\sim$ 7.0 mJ/mol$\cdot$K$^2$ is observed in the specific heat $C(T)$, as shown in the inset of Fig. 1(b).  If we subtract the core diamagnetism $\chi_{\rm core} = -0.8$ $\times$ 10$^{-4}$ emu/mol \cite{Bain2008} from $\chi_0$ and assume $\chi_{0} - \chi_{\rm core}$ = 2.3 $\times$ 10$^{-4}$ emu/mol represents the Pauli paramagnetic susceptibility $\chi_{\rm P}$, the observed $\gamma$ and the naively estimated $\chi_{\rm P}$ yield a Wilson ratio $R_{\rm W}$ = 2.4. The estimated $R_{\rm W}$ is appreciably larger than the unity for non-interacting electrons, which at a glance would imply strongly correlated nature of electrons and possibly additional Stoner enhancement. We note however that $\chi_0 - \chi_{\rm core}\;$should include not only $\chi_{\rm P}$ but also van Vleck contribution$\;\chi_{\rm VV}$, which can be as large as $\sim10^{-4}$ emu/mol in iridates \cite{Chaloupka2013}. If it is the case, the Wilson ratio $R_{\rm W}$ is very close to 1 rather than the naïve estimate using $\chi_0 - \chi_{\rm core}$. This gives a reasonable doubt on the enhanced $R_{\rm W}$ in the monoclinic SrIrO$_3$.

\begin{figure}[htb]
\begin{center}
\includegraphics*[scale=0.15]{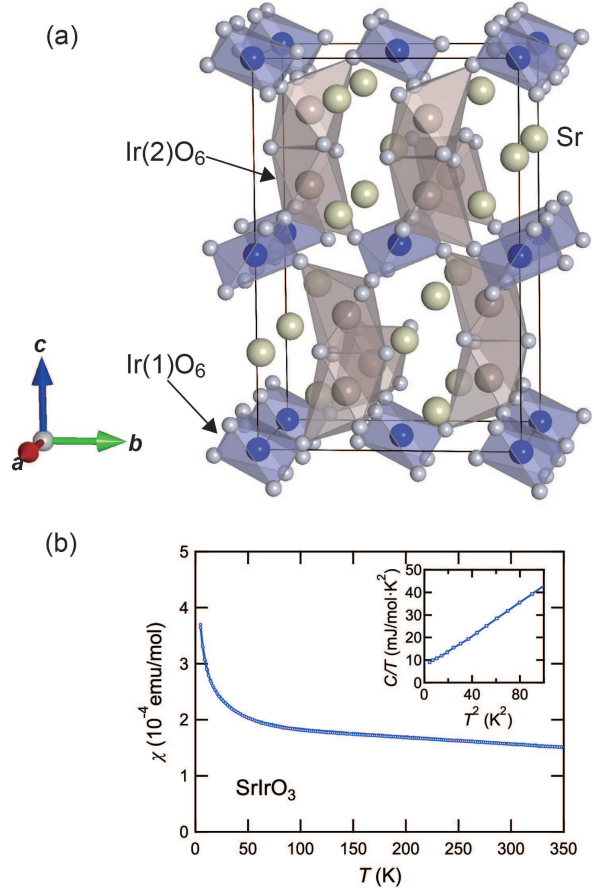}
\caption{(a) Crystal structure of monoclinic SrIrO$_3$ \cite{Momma2008}. The blue and brown octahedra depict Ir(1)O$_6$ and Ir(2)O$_6$, respectively. (b) Temperature dependence of magnetic susceptibility $\chi(T)$. The inset shows the specific heat $C(T)$ divided by temperature with respect to the squared temperature.}
\end{center}
\end{figure}

The residual resistivity of $\rho_{0} \sim$ 0.5 m$\Omega$cm is large as a metal even if we consider that the sample is a polycrystalline pellet, indicating that the system is poorly metallic. Nevertheless, the magnetoresistance $\Delta\rho/\rho_{0} = (\rho(B) - \rho_{0})/\rho_{0}$ at 5 K, shown in the inset of Fig. 2(a), is as large as 4\% at a magnetic field $B$ of 9 T, indicative of a moderately high mobility $\mu$ of charge carriers ($\mu \sim$ 200 cm$^2$/V$\cdot$s). This likely suggests that the poorly metallic transport originates from a low carrier density, not from the strong disorder. The Hall constant shown in Fig. 2(b) indeed evidences the low density of carriers. The $R_{\rm H} (T)$ displays a pronounced temperature dependence and a large absolute value at low temperatures, $\mid R_{\rm H}(5\;{\rm K})\mid\; \sim$ 0.014 cm$^3$/C . This corresponds to the nominal carrier number $n$ of $\sim4.5 \times 10^{20}$ cm$^{-3}$, which is two orders of magnitude smaller than that of half-filled $J_{\rm eff}$ = 1/2 metal picture ($n \sim 1.5 \times$ 10$^{22}$ cm$^{-3}$ if one electron per iridium atom is assumed).

The Seebeck constant $S(T)$ also shows a complicated temperature dependence. $S(T)$ increases on cooling from room temperature, displays a broad peak at around 50 K and then decreases at low temperatures. While $R_{\rm H}(T)$ is negative below room temperature, $S(T)$ shows a positive value in the whole measured temperature range, suggesting that both holes and electrons are present and therefore the system is semimetallic. The magnitude of Nernst coefficient $\nu(T)$ is about $\sim0.2$ $\mu$V/KT at low temperatures, which is quite large for a metal and suggests an ambipolar effect with the presence of two-types of carriers \cite{Behnia2009}. These facts clearly point to the semimetallic electronic structure of monoclinic SrIrO$_3$.

\begin{figure}[htb]
\begin{center}
\includegraphics*[scale=0.2]{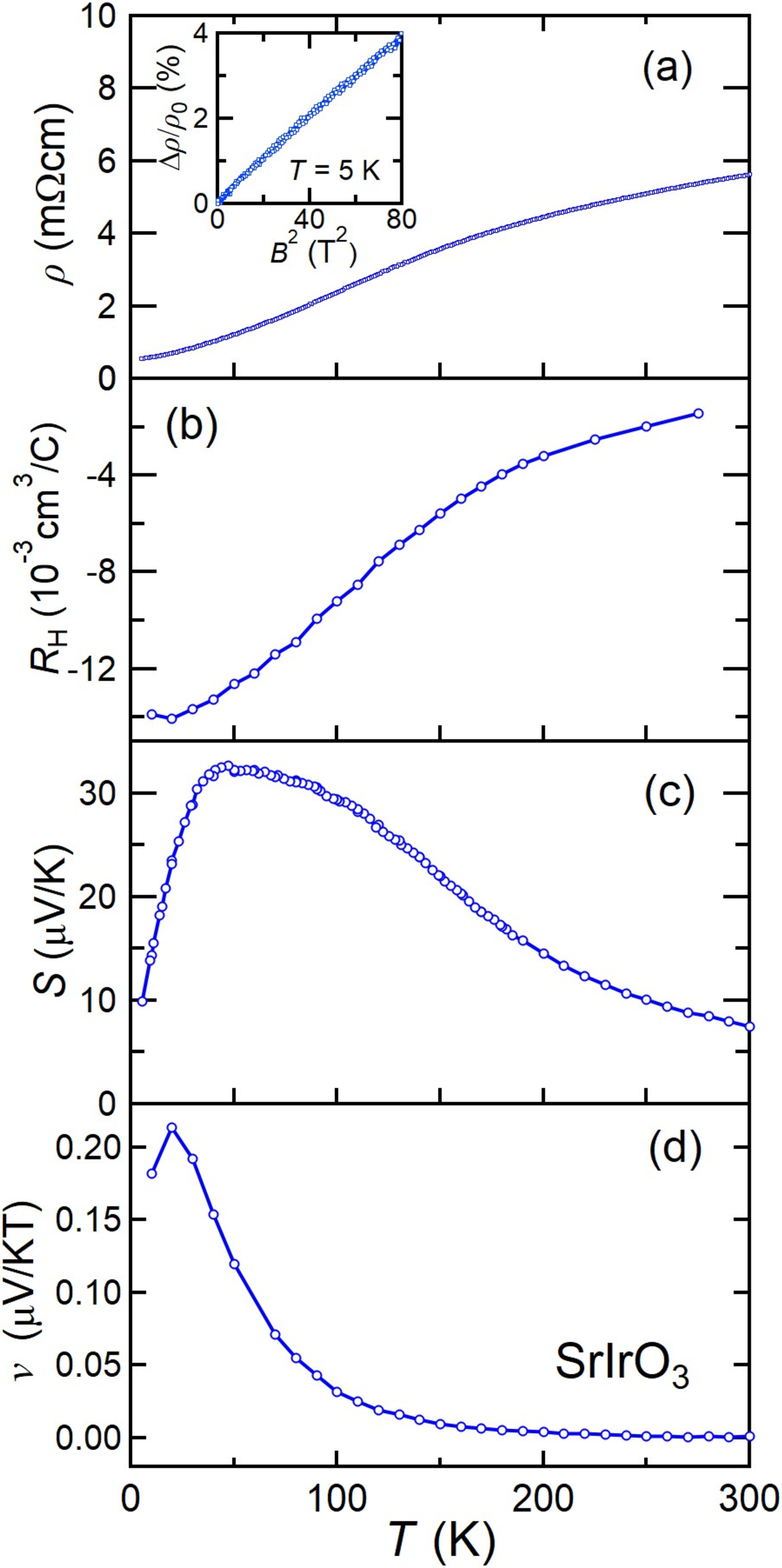}
\caption{Transport properties of monoclinic SrIrO$_3$ displaying the temperature dependences of (a) resistivity $\rho(T)$, (b) Hall constant $R_{\rm H}(T)$, (c) Seebeck constant $S(T)$, and (d) Nernst coefficient $\nu(T)$. The inset of (a) shows the transverse magnetoresistance $\Delta\rho/\rho_{0}$ ($\Delta\rho = \rho(B) - \rho_{0}$, and $\rho_{0}$ is the resistivity at $B$ = 0 T) at 5 K.}
\end{center}
\end{figure}

\subsection{Electronic structure calculation}

In order to elucidate the origin of semimetallic state of monoclinic SrIrO$_3$, we performed electronic structure calculations for the experimental crystal structure parameters. Figure 3(a) shows the band dispersions near the Fermi energy $E_{\rm F}$ along the high-symmetry lines (see the Brillouin zone in Fig.3(c)), which is calculated without spin-orbit coupling. The density of states (DOS) of $t_{2g}$ states shown in Fig.3(b) is resolved into the contributions from the Ir(1) and Ir(2) sites, which are found to be almost identical in the energy region near $E_{\rm F}$. There is a broad DOS peak around $E_{\rm F}$, incompatible with the semimetallic transport properties. The formation of molecular orbitals of $a_{1g}$ orbitals in the face-shared Ir$_2$O$_9$ unit was discussed in some iridium oxides where the two $a_{1g}$ orbitals pointing towards the face-shared bond overlap substantially \cite{Streltsov2017, Ye2018}. The clear split of bonding- and antibonding $a_{1g}$ orbitals, however, is not identified in the density of states for Ir(2) sites, suggesting the strong hybridization of the $t_{2g}$ states between Ir(1) and Ir(2) sites.

The incorporation of spin-orbit coupling has a strong impact on the electronic structure. Figure 4 shows so-called fat bands and densities of $d$ states for Ir(1) and Ir(2) atoms resolved into their $J$ characters, $J$ = 5/2 ($J_{5/2}$) and $J$ = 3/2 ($J_{3/2}$). It should be noted that the $J_{\rm eff}$ = 1/2 wave function consists purely of $J$ = 5/2 states while $J_{\rm eff}$ = 3/2 contains both $J$ = 5/2 and 3/2 components. As in the calculation without spin-orbit coupling (Fig. 3), the DOS for the Ir(1) and Ir(2) sites are almost identical. For both Ir sites, the conduction bands have primarily $J$ = 5/2 character indicating the dominant contribution of the $J_{\rm eff}$ = 1/2 state. On the other hand, both $J$ = 5/2 and $J$ = 3/2 characters are seen below $E_{\rm F}$, pointing to the strong hybridization of $J_{\rm eff}$ = 1/2 and 3/2 states. A gap is about to open and a steep dip is observed in the DOS for both Ir(1) and Ir(2) sites at $E_{\rm F}$. The small DOS seems to be consistent with the semimetallic state. With a close inspection of the states near $E_{\rm F}$ shown in Fig. 4(e), we indeed find a small overlap of the conduction band and the valence bands, giving rise to an electron pockets at the A point and a hole pocket at the M point of Brillouin zone (Fig. 4(f)). The valence and conduction bands forming the two pockets around the A and M points have a crossing slightly below and above $E_{\rm F}$ respectively. The crossings represent a Dirac point. The monoclinic SrIrO$_3$ is thus found to be a semimetal with two protected gapless Dirac points near $E_{\rm F}$.

\begin{figure}[htb]
\begin{center}
\includegraphics*[scale=0.17]{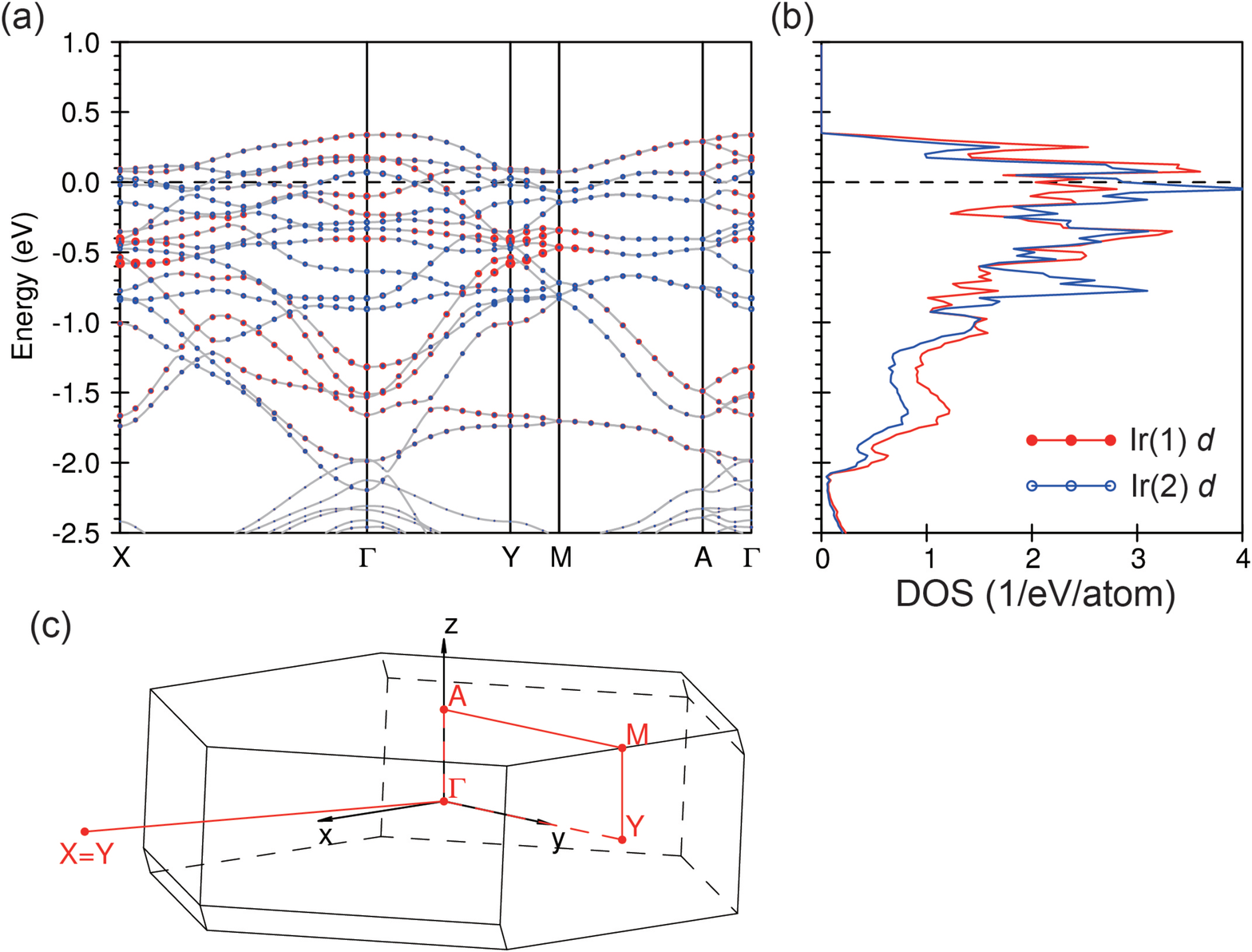}
\caption{(a) Band structure and (b) density of Ir $d$ states of monoclinic SrIrO$_3$ calculated without including spin-orbit coupling. The density of states is resolved into the contributions of $d$-orbitals from the Ir(1) and Ir(2) sites. The size of red and blue circles in (a) represents the weight of $d$-orbitals from Ir(1) and Ir(2) sites, respectively. (c) the Brillouin zone of $C2/c$ monoclinic structure.}
\end{center}
\end{figure}

\section{Discussion}

The result of band calculation indicates that the semimetallic electronic structure of monoclinic SrIrO$_3$ originates from the presence of two protected (gapless) Dirac points near $E_{\rm F}$. Since the crystal structure can be viewed as a stack of buckled honeycomb bilayers of IrO$_6$ octahedra through the face-sharing connection of Ir(2)$_2$O$_9$ unit, the relevance to the Dirac bands on a honeycomb lattice might be expected. However, the Dirac band of buckled honeycomb lattice is gapped in the presence of trigonal distortion and is present only at the corner of two-dimensional Brillouin zone \cite{Xiao2011, Lado2013}. These behaviors seem incompatible with the gapless Dirac dispersion at the A and M points in the calculated band structure of monoclinic SrIrO$_3$. We argue that the Dirac points are protected by its non-symmorphic symmetry (space group $C2/c$) \cite{Gibson2015, Young2015, Schoop2016, Chen2017}. Because of the presence of $c$-glide symmetry, the bands in the $k_{z}$ = $\pi$ plane do not hybridize with each other when they possess different glide-eigenvalues \cite{Fang2015}. Indeed, we see the four-fold degeneracy (i.e. crossing of two Kramers pairs) at the A and M points in the whole energy region displayed in Fig. 4.

\begin{figure}[htb]
\begin{center}
\includegraphics*[scale=0.18]{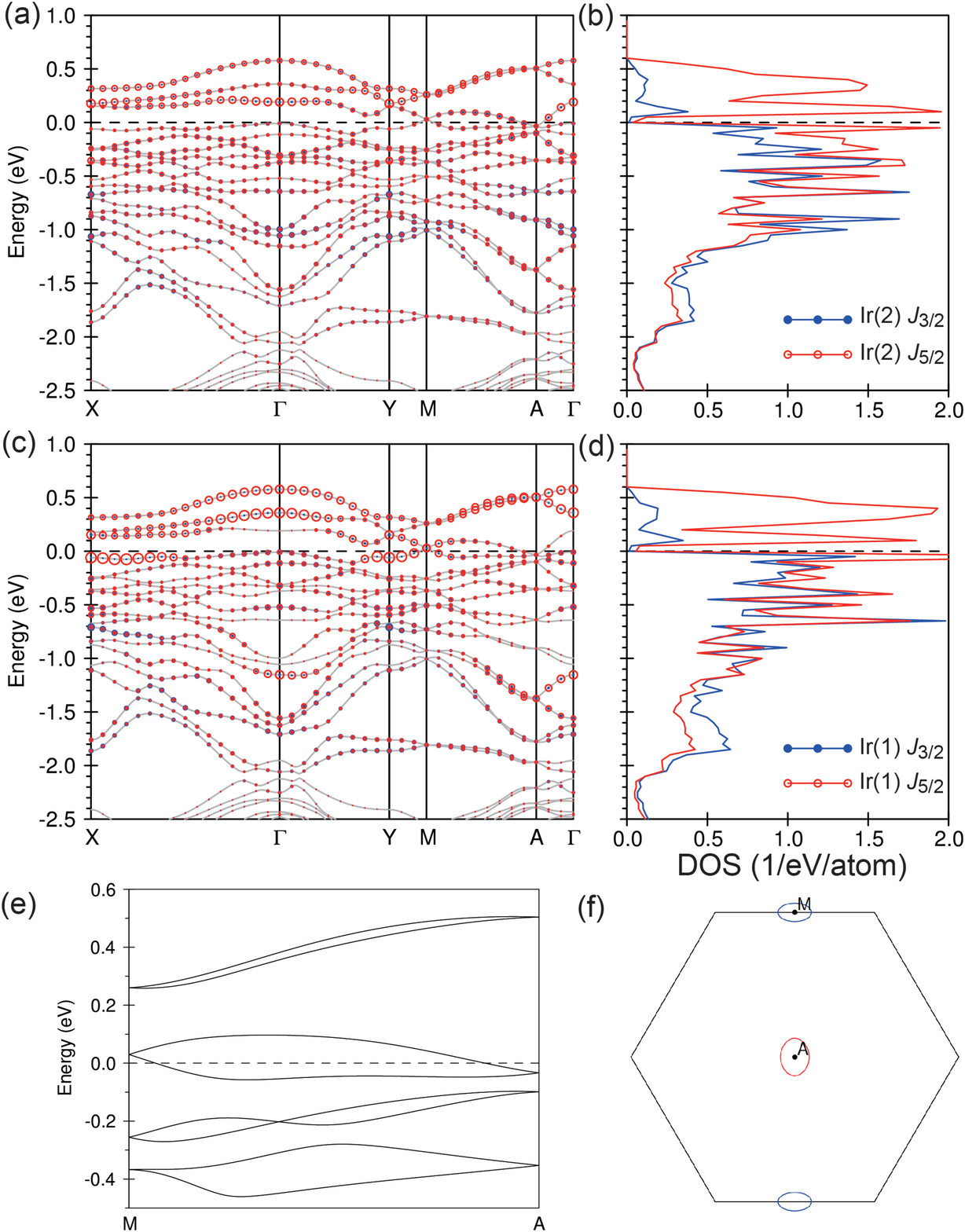}
\caption{(a), (c) the fat band structure and (b), (d) densities of Ir(2) and Ir(1) $d$ states of monoclinic SrIrO$_3$ calculated including spin-orbit coupling. In (b), (d), the density of states are resolved into $J$ = 3/2 and 5/2 characters for each site. The size of blue and red circles in (a) and (c) represents the weights of $d$-orbital characters with $J$ = 3/2 and 5/2, respectively. (e) The magnified view of bands near $E_{F}$ at the M and A point highlighting the Dirac bands. (f) Fermi surface cross section in the $k_{z} = \pi$ plane of Brillouin zone.}
\end{center}
\end{figure}

The symmetry-protected degeneracy at the A and M points does not depend on whether spin-orbit coupling is incorporated or not. The calculation without spin-orbit coupling in Fig. 3 indeed shows the degenerate bands at these points. In the absence of spin-orbit coupling, the bands meeting at the A and M points remain degenerate along the A-M line. These bands, together with other bands crossing the $E_{\rm F}$ along the $\Gamma$-X and $\Gamma$-Y lines, give rise to the large DOS at $E_{\rm F}$. By turning on the spin-orbit coupling, the bands along the $\Gamma$ -X and $\Gamma$-Y lines are pushed away from $E_{\rm F}$. Furthermore, the degeneracy along the A-M line is lifted, whereas the crossings at the A and M points, protected by the non-symmorphic symmetry, remain robust against spin-orbit coupling. The split of bands by spin-orbit coupling gives rise to the Dirac points at the A and M point and renders small pockets Fermi surfaces.

Chen et al. discussed the filling condition for the appearance of semimetallic state in non-symmorphic materials \cite{Chen2017}. They proposed that the electron count in formula unit v$_{\rm F}$ and $Z$* = $Z/W_{a}^{\rm G}$ need to be odd to realize a filling-enforced semimetal where $Z$ is the number of formula unit in the primitive cell and $W_{a}^{\rm G}$ is the multiplicity of the highest-symmetry Wyckoff position of the space group G (measured with respect to the primitive cell), respectively. The monoclinic SrIrO$_3$ indeed satisfies the condition as v$_{\rm F}$ = 139 and $Z$* = 6/2 = 3.
 
\section{Conclusion}
We found that the monoclinic SrIrO$_3$ is a semimetal from the transport measurements. The electronic structure calculations showed that the semimetallic state is composed of the two protected Dirac points below $E_{\rm F}$ at the A and above $E_{\rm F}$ at the M points of Brillouin zone, respectively. We argue that the Dirac bands are produced by the non-symmorphic symmetry in combination with the strong spin-orbit coupling. The polymorphic SrIrO$_3$ with the GdFeO$_3$-type perovskite structure has been recently attracting attentions as a topological semimetal produced by strong spin-orbit coupling, and the emergence of a variety of topologically nontrivial phases is anticipated by fine tuning of lattice symmetries \cite{Chen2015, Fang2016}. We expect that monoclinic SrIrO$_3$ can be another platform for nontrivial topological phases driven by strong spin-orbit coupling.

\ack
We thank K. Ohashi for experimental support and discussions in the early stage of this study. This work was partly supported by the Alexander von Humboldt foundation and Japan Society for the Promotion of Science (JSPS) KAKENHI (No. JP15H05852, JP15K21717, 17H01140).

\section*{References}
\bibliography{references_SrIrO3}

\end{document}